\documentclass[conference]{IEEEtran}\IEEEoverridecommandlockouts
\usepackage{etoolbox}
\usepackage{algorithmic}
\usepackage{algorithm}

\makeatletter
\patchcmd{\@makecaption}
  {\scshape}
  {}
  {}
  {}
\makeatletter
\patchcmd{\@makecaption}
  {\\}
  {.\ }
  {}
  {}
\makeatother
 \usepackage{amsmath,amssymb}
 \usepackage{subfigure}
 \usepackage{graphicx, graphics,color,psfrag}
 \usepackage{cite,balance}
 \usepackage{caption}
 \allowdisplaybreaks
 \usepackage{algorithm}
 \usepackage{accents}
 \usepackage{amsthm}
 \usepackage{bm}
 \usepackage{algorithmic}
 \usepackage[english]{babel}
 \usepackage{multirow}
 \usepackage{enumerate}
 
 \usepackage[T1]{fontenc}
 \usepackage{aecompl}

 \usepackage{cases}
 \usepackage{stfloats}
 \usepackage{dsfont}
 \usepackage{color,soul}
 \usepackage{amsfonts}
 \usepackage{cite,graphicx,amsmath,amssymb}
 \usepackage{subfigure}
 \usepackage{fancyhdr}
 \usepackage{hhline}
 \usepackage{graphicx,graphics}
 \usepackage{array,color}
 \usepackage{amsmath}
\usepackage{float}
\usepackage{amssymb}
\usepackage{amsmath}
\usepackage{amsthm}
\usepackage{amsfonts}
\usepackage{graphicx}

\usepackage{epstopdf}
\usepackage{cite}
\usepackage{amsmath,bm}
\usepackage{subfigure}
\usepackage{graphicx}
\usepackage{color}
\usepackage{graphicx}
\usepackage{calc}
\usepackage{caption}

\newtheorem{remark}{\textbf{Remark}}

\begin{document}

\title{Two Birds with One Stone: Multi-Task Semantic Communications Systems over Relay Channel}
\author{Yujie Cao, Tong Wu, Zhiyong Chen, Yin Xu, Meixia Tao, Wenjun Zhang\\
		Cooperative Medianet Innovation Center, Shanghai Jiao Tong University, Shanghai, China\\
		Email: \{1452377525, wu\textunderscore tong, zhiyongchen, xuyin, mxtao, zhangwenjun\}@sjtu.edu.cn}
\maketitle
\begin{abstract}
  In this paper, we propose a novel multi-task, multi-link relay semantic communications (MTML-RSC) scheme that enables the destination node to simultaneously perform image reconstruction and classification with one transmission from the source node. In the MTML-RSC scheme, the source node broadcasts a signal using semantic communications, and the relay node forwards the signal to the destination. We analyze the coupling relationship between the two tasks and the two links (source-to-relay and source-to-destination) and design a semantic-focused forward method for the relay node, where it selectively forwards only the semantics of the relevant class while ignoring others. At the destination, the node combines signals from both the source node and the relay node to perform classification, and then uses the classification result to assist in decoding the signal from the relay node for image reconstructing. Experimental results demonstrate that the proposed MTML-RSC scheme achieves significant performance gains, e.g., $1.73$ dB improvement in peak-signal-to-noise ratio (PSNR) for image reconstruction and increasing the accuracy from $64.89\%$ to $70.31\%$ for classification.
\end{abstract}

\section{Introduction}
With the rapid advancement of artificial intelligence (AI), semantic communications have recently garnered significant attention, as they represent the second level of information transmission outlined by Shannon and Weaver \cite{Shannon}. This paradigm shift in communications moves the focus from the accurately conveying symbols to effectively transmitting the intended meaning. By employing AI to extract and transmit the semantic essence of the source data, semantic communications significantly enhance the efficiency of the communication process. Consequently, semantic communications are regarded as a promising approach to fulfilling the demanding specifications of sixth-generation (6G) wireless networks \cite{CDDM}.

Previous studies have explored various scenarios in semantic communications \cite{DBC,gundu2019,SwinJSCC,mambajscc}. Three key works focus on the foundational design of joint source-channel coding (JSCC) for semantic communications. In \cite{gundu2019}, DeepJSCC is introduced for image transmission based on convolutional neural networks (CNNs), aiming to optimize performance. An enhanced version, SwinJSCC, is later proposed in \cite{SwinJSCC}, incorporating advanced Transformer modules. More recently, MambaJSCC, a lightweight and efficient deep JSCC architecture for wireless image transmission, was developed in \cite{mambajscc} based on  selective structure state space models (SSM). Building on these works, research on multi-task semantic communications has advanced in \cite{Tao, Zhangping, Qin}. In \cite{Tao}, additional decoder heads are introduced to handle different tasks, while \cite{Zhangping} proposes a Feature Importance Ranking module to assess the impact of each feature on task performance. Moreover, \cite{Qin} divides the features into shared and private groups for each task, allowing the node to selectively transmit features, thereby improving efficiency.

In practice, communication via a relay node is common, as direct communication between distant parties often suffers from poor signal quality. Unlike the traditional amplify-and-forward (AF) and decode-and-forward (DF) methods, \cite{SF} introduces the semantic-and-forward (SF) approach for semantic communications, where the relay first interprets and then forwards semantic information. This approach has been further developed in works such as \cite{SLF, SFR, RSC, HEC, SemF}. Due to the broadcasting nature of the source node’s transmission, the signal can also reach the destination node through the direct source-destination (SD) link. Although the SD link typically has lower quality, it still provides valuable information that can enhance performance, for example, by using traditional maximum ratio combining (MAP) techniques, which are often overlooked in existing studies. Only one study \cite{SD} explores the use of the SD link in semantic relay communication, where the destination node employs a semantic combining method to integrate signals from both the SD link and the relay-destination (RD) link. However, this study focuses solely on text transmission and lacks a specialized design for multi-task communication.

Motivated by this, we propose a multi-task, multi-link relay communications (MTML-RSC) scheme that enables the destination node to simultaneously perform image reconstruction and classification with a single transmission from the source node. We analyze the characteristics of the two tasks, noting that classification requires significantly less information than reconstruction, allowing it to be completed with high quality via the direct link. Additionally, the classification result can be served as side information, guiding and improving the reconstruction process. Therefore, we propose a semantic-focused forward mode at the relay node in the proposed MTML-RSC scheme, where the relay extracts and forwards the semantics of specific classes while disregarding the semantics of irrelevant classes. Furthermore, the improved reconstruction can also enhance classification by providing more accurate semantic information. Based on this, the destination node combines signals from both the source and relay nodes to perform classification, then leverages the classification result to aid in decoding the relay signal for image reconstructing. Experimental results show that the proposed MTML-RSC scheme delivers notable performance improvements, e.g., a $1.73$ dB boost in peak signal-to-noise ratio (PSNR) for image reconstruction and a rise in classification accuracy from $64.89\%$ to $70.31\%$.

\section{System Model}\label{I}
In this section, we propose our MTML-RSC scheme for both image reconstruction and classification under the relay channel, combining the direct link. 
\subsection{System overview}

\begin{figure*}[t]
  \begin{center}
    \includegraphics[width=\textwidth]{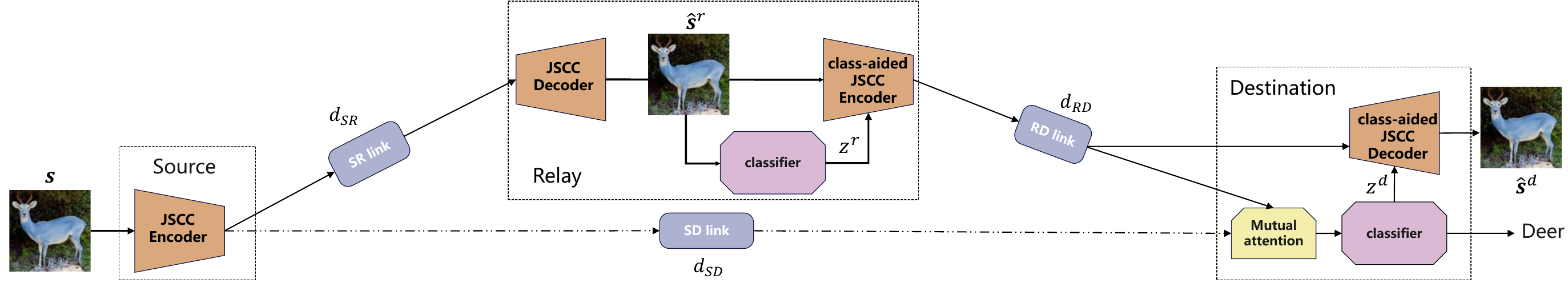}
  \end{center}
    \caption{The overall architecture of the proposed MTML-RSC scheme.}
    \label{system_model}
    \vspace{-0.3 cm}
\end{figure*}

Our scheme consists of three nodes: the source node, the relay node, and the destination node as shown in Fig. \ref{system_model}. 
\subsubsection{Source node}
The source node is composed of a JSCC encoder to extract and encode the source image $\mathbf{s} \in \mathbb{R}^{3\times H\times W}$ to the channel input signal $\mathbf{x}^s\in \mathbb{C}^{n \times \frac{l}{2}}$ with power constraint $\mathbb{E}||\mathbf{x}^s||_2^2 \leq P_s$, where the $H$, $W$ is the height and weight of the image and $3$ represents the color channel R, G and B. $l$ denotes the total length of the real and imaginary part of the complex channel input signal and $n$ denotes the signal patches. 
The JSCC encoder model consists of two stages with Swin Transformer as its backbone modules. In the first stage, the source image is transformed to $c_1^e$ learnable patches with size $h_1^e\times w_1^e$ by the patch embedding module and then $n_1^e$ Swin Transformer modules are applied to these patches to furtherly learn the effective representation of these patch with their number and size unchanged. The output patches are then fed into the second stage, comprising a downsampling patch merging module, outputting $c_2^e$ patches with size $h_2^e\times w_2^e$, and additional $n_2^e$ Swin Transformer modules. Finally, a fully connected (FC) module is applied on these patches, followed by power normalization for the channel input signal. 
The signal $\mathbf{x}^s$ is then broadcasted to both the relay node and the destination node under the source-raley (SR) and SD links. 

\subsubsection{Relay node}
The relay node applies our proposed semantic-focused forward, which is composed of a JSCC decoder, a classifier, and a classification-aided JSCC encoder. The JSCC decoder reconstructs the image as $\mathbf{\hat{s}}^r\in \mathbb{R}^{3\times H\times W}$ from $\mathbf{y}^r$, followed by the classifier performing label $z^r$ to the reconstructed image. After that, a classification-aided JSCC encoder takes the $\mathbf{\hat{s}}^r$, along with  $z^r$, as input to recode the image as $\mathbf{x}^r \in \mathbb{C}^{n \times \frac{l}{2}}$ for forwarding to the destination through RD link with power constraint $\mathbb{E}||\mathbf{x}^r||_2^2 \leq P_r$. 

The JSCC decoder has a symmetrical structure to the encoder, which also consists of two stages with Swin Transformer as its backbone module. Firstly, an FC module is applied to expand the received signal to $c_2^d$ patches with size $h_2^d\times w_2^d$. The output patches are then fed into stage 2, which comprises $n_2^d$ Swin Transformer modules and an upsampling patch division module, outputting $c_1^d$ patches with size $h_1^d\times w_1^d$. In stage 1, $n_1^d$ Swin Transformer modules are applied to these patches to further learn the effective representation of these patches with their number and size unchanged. These patches are then passed through an upsampling patch division module, followed by a reshaping process to reconstruct the source image.

\subsubsection{Destination node}
The destination node applies the proposed multi-link combining receiving, which is composed of a classification-aided JSCC decoder, a classifier, and a mutual attention module. Firstly, the mutual attention module integrates $\mathbf{y}^r$ and $\mathbf{y}^f$ from the two links into a fused signal $\mathbf{y}$, followed by the classifier performing label $z^d$ from $\mathbf{y}$. After that, a classification-aided JSCC decoder takes the $\mathbf{y}^f$ and $z^d$ as input to reconstruct the source image as $\mathbf{\hat{s}}^d\in \mathbb{R}^{3\times H\times W}$.

\subsection{Channel model}
Typically, we consider the links to be Rayleigh fading with different SNR and large-scale fading. Therefore, the $i$-th element of received signals at the relay node $y_i^r$ and destination node $y_i^d$ obtained from the SR link and the SD link can be expressed respectively as
\begin{align}\label{SR-SD}
   y_i^r=d_{SR}^{-a}h^r_ix^s_i+n^r,\\
   y_i^d=d_{SD}^{-a}h^d_ix^s_i+n^d,
\end{align}
where $h^r_i$, $h^d_i \sim \mathbb{CN}(0,1)$ are independent and identically distributed (i.i.d.) Rayleigh fading gains, $d_{SR}$ and $d_{SD}$ are the distance of the SR link and SD link, $n^r$ and $n^d$ are i.i.d. Additive White Gaussian Noise (AWGN) samples with variance $\sigma^2_r$ and $\sigma^2_d$, and $a$ is the large-scale fading exponent.
Here we set $d_{SR}<d_{SD}$ and $\sigma_r \leq \sigma_d$, reflecting the typical relay scenario that the channel condition the SR link is superior to that of the SD link.

Analogous to the preceding links, the $i$-th element of received signal $y_i^f$ from the RD link can be expressed as
\begin{align}\label{receive signal}
y_i^f=d_{RD}^{-a}h^f_ix^r_i+{n^f_i},
\end{align}
where ${h^f_i} \sim \mathbb{CN}(0,1)$ are i.i.d Rayleigh fading gain of RD link, $d_{RD}$ is distance between them, $n_i^f$ is i.i.d AWGN sample with variance $\sigma^2_f$.
The RD link is also better than the SD link as the common scenario in relay communications such that $\sigma_f \leq \sigma_r$ and $d_{RD}<d_{SD}$.
We also consider the AWGN channel with large-scale fading, where the Rayleigh fading gains $h^r_i, h^d_i, h^f_i$ are ignored as constant $1$.
\begin{figure}[t]
  \begin{center}
    \includegraphics[width=0.48\textwidth]{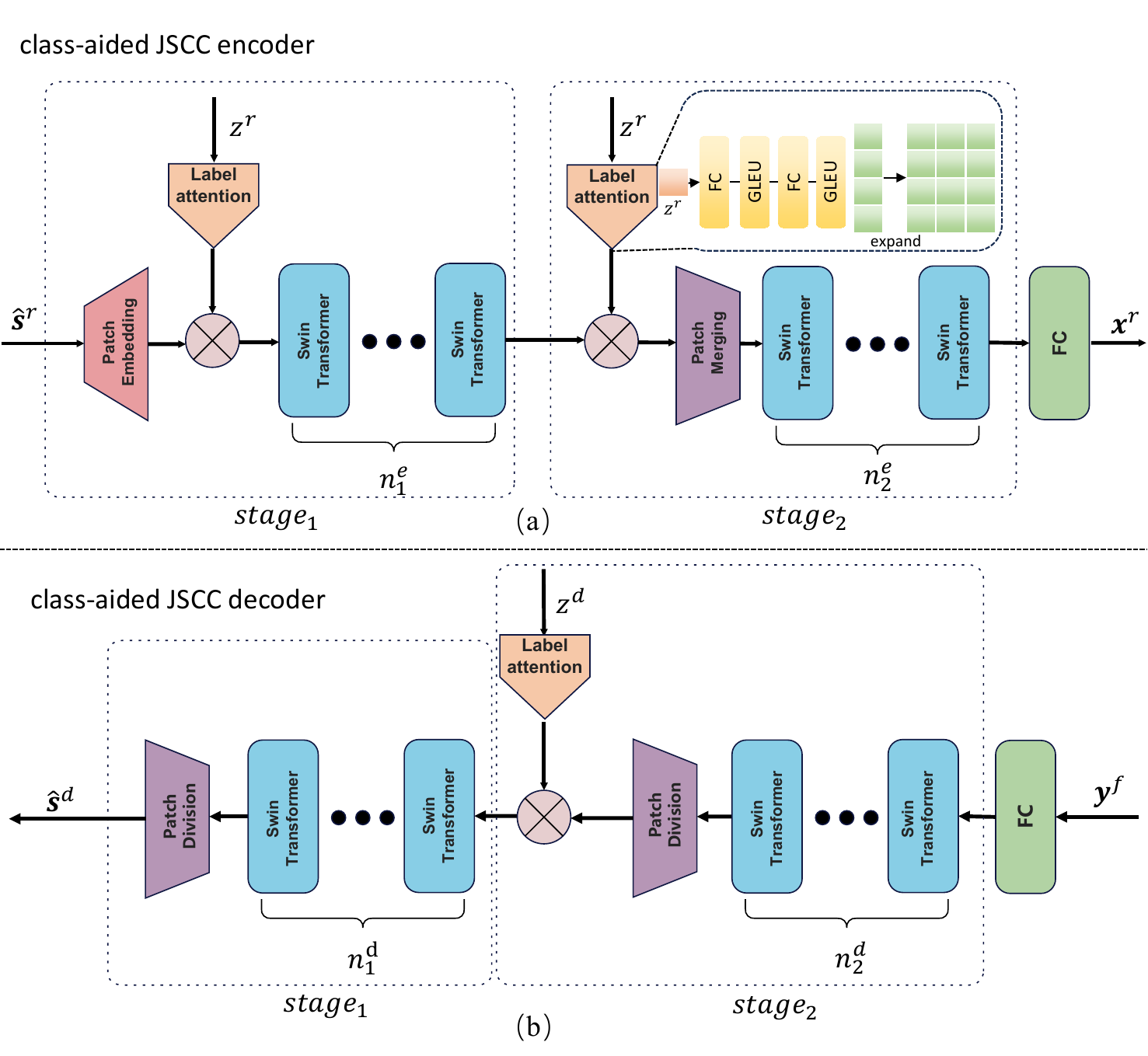}
  \end{center}
    \caption{The structure of the designed class-aided JSCC encoder and decoder.}
    \label{encoder-class}
    \vspace{-0.3 cm}
\end{figure}

\section{Relay and receiver design}
In this section, we provide a detailed illustration of the proposed semantic-focused forward and the multi-link combining receiving technologies.
\subsection{Semantic-focused forward}
As previously discussed, the destination node receives both the forwarded signal and the broadcasted signal through the RD and SD links. Although the signal experiencing from SD link is severely degraded compared to that from the RD link, the destination node can still understand some semantic information from it. Therefore, the relay node is required to selectively forward the semantic information that remains ambiguous to the destination node only from SD link to utilize the RD link more effectively.
However, this poses a challenge, as the relay node has no knowledge of the semantics which the destination node has understood through the SD link. To address the problem, we first analyze the characteristics of the classification and reconstruction tasks to provide prior knowledge for the relay node.

On one hand, the two tasks are interrelated, as better reconstruction includes the better recovery of class features, leading to more accurate classification, and accurate classification can also guide to understand the semantic for better reconstruction. On the other hand, the two tasks require distinct amounts of information, with accurate classification needing significantly less information than high-quality reconstruction. Based on the observation, we assume that the destination node can understand the classification-related semantics through the degraded SD link. Thus, the relay node can concentrate on the semantics of relevant class, preserving transmission capacity of the RD link for reconstruction.

Based on the above analysis, we propose a semantic-focused forward relay mode, where the relay node first decodes the received signal for reconstruction, followed by a classifier recognizing the image class. After that, the class-aided JSCC encoder focuses on re-encoding and forwarding the specific semantics of this class, while the semantics of other classes are ignored. Thus, in the semantic-focused forward, only the most relevant semantics to the class are forwarded, while the semantic of other class are ignored, thus enhancing coding efficiency for enhanced performance.
\begin{figure}[t]
  \begin{center}
    \includegraphics[width=0.49\textwidth]{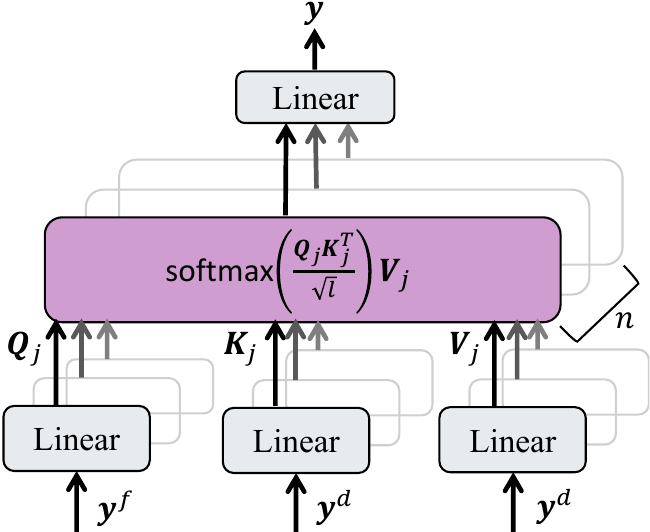}
  \end{center}
    \caption{The structure of the designed mutual attention module.}
    \label{c-attention} 
    \vspace{-0.3 cm}
\end{figure}

\begin{remark}
  The improvement in the transmission quality and performance can be understood through the perspective of entropy. It is well known that in our scenario, the entropy of the reconstructed image at the relay node under the condition of its class is always less or equal to its entropy, which can be expressed as
\begin{align}\label{conditional entropy}
  H(\mathbf{\hat{s}}^r\mid z^r) \leq H(\mathbf{\hat{s}^r}).
  \end{align}
Therefore, in the semantic-focused forward mode, the information expected to be coded and forwarded is reduced. The reduced part is acquired through the SD link, thus avoiding redundant information transmission for enhancing the transmission efficiency.
\end{remark}

The semantic-focused forward is achieved by the classifier and the designed class-aided JSCC encoder. The classifier takes the reconstructed image $\mathbf{\hat{s}}^r$ as input and extracts it feature through the similar architecture of the JSCC encoder, except that the mean of the output from the final stage is processed by a multi-layer perceptron (MLP) for classification. The class-aided JSCC encoder also is built with two stage with Swin Transformer, as shown in Fig. \ref{encoder-class}(a).
To be specific, at the end of each stage, we design a label attention module to inject class-related information into the encoder, where the module learns to map the class label $z^r$ into a sequence of the same size as the patches through several FC layers and activation function $GLEU(\cdot)$, then expands it to match the number as the patches. Then, the expanded sequences are multiplied with the patches in an element-wise manner. Therefore, the label attention module can dynamically adjust the weight of each patch element, guiding the subsequent blocks in the encoder to focus on the semantics of this class.



\addtolength{\topmargin}{0.05in}
\subsection{Multi-link combining receiving}

In cooperation with the semantic-focused forward, the destination node is required to recognize the class from the SD link. Moreover, based on the analysis that the two tasks are benefitial to each other. Therefore, it is better to combine the signals from both links for classification. On the other hand, the source image should be reconstructed using the signal from the RD link, with class information acquired from the classification task. Thus, the reconstruction task is completed after the classification task for combining the classification result.
Based on this analysis, we propose a multi-link combining receiving, 
where the signal from both SD and RD links are received as real number with $n$ patches of length $l$ and combined by the designed mutual attention module, which maps the $n$ signal groups with size $l$ from the SD link to $\mathbf{K}_j$, $\mathbf{V}_j$, $j=1,2,...,n$ by $2n$ FC modules and the $n$ signal groups from the RD link are projected to $\mathbf{Q}_j$ by another $n$ FC modules. Then the softmax multi-head attention is adopted to them as
\begin{align}
  \mathbf{y}^j = \text{softmax}(\frac{\mathbf{Q}_j\mathbf{K}_j^T}{\sqrt{l_s}})\mathbf{V}_j.
\end{align}
These $n$ outputs are then concentrated and addressed by another FC as final output $\mathbf{y}\in \mathbb{R}^{n\times l}$. The structure of the mutual attention module is shown in Fig. \ref{c-attention}.
$\mathbf{y}$ is then fed into the classifier for classification. 
After classification, the class-aided JSCC decoder combined the classifier output $z^d$ and the signal from the RD link to reconstruct the source image. The class-aided JSCC decoder consist of the symmetrical architecture of the class-aided JSCC encoder as shown in Fig. \ref{encoder-class}(b), while the label attention module only exists at the end of the second stage.
The combined reconstruction process ensures that the decoder is synchronized with the encoder, allowing them to focus on the same class for improving reconstruction quality.
\begin{algorithm}[t]
  \hspace*{0.02in} {\bf \small{Input:}}
	\small{Training image distribution $p(\mathbf{s})$, image class distribution $p_\mathbf{s}(z)$ and the channel estimation result $\mathbf{h}^r$, $\mathbf{h}^d$, $\mathbf{h}^f$.} \\
	\hspace*{0.02in} {\bf \small{Output:}}
	\small{The trained joint new architecture.}
	\caption{Training algorithm of proposed MTML-RSC scheme.}
	\label{trainCDESC}
	\begin{algorithmic}[1]
    \WHILE {the training stop condition of stage one is not met }
    \STATE Samply $\mathbf{s}$ from $S$ randomly. 
    \STATE Encode $\mathbf{s}$ as $\mathbf{x}^s$ by the JSCC encoder.
    \STATE Transmit $\mathbf{x}^s$ through SR link as $\mathbf{y}^r$ following (\ref{SR-SD}).
    \STATE Decode $\mathbf{y}^r$ by the JSCC decoder for reconstructed as $\mathbf{\hat{s}}^r$.
    \STATE Compute $L_1$ loss and update the parameters of the JSCC encoder and the JSCC decoder.
    \ENDWHILE
    \WHILE {the training stop condition of stage two is not met }
    \STATE Samply $\mathbf{s}$ from $S$ randomly.
    \STATE Predict the class $z^r$ of $\mathbf{\hat{s}}^r$ by the classifier at the relay node.
    \STATE Compute $L_2$ loss and update the parameters of the classifier.
    \ENDWHILE
    \WHILE {the training stop condition of stage three is not met }
    \STATE Samply $\mathbf{s}$ from $S$ randomly.
    \STATE Encode $\mathbf{\hat{s}}^r$ as $\mathbf{x}^r$ by the class-aided JSCC encoder with its true label $z$.
    \STATE Transmit $\mathbf{x}^r$ through RD link as $\mathbf{y}^f$ following (\ref{receive signal}).
    \STATE Receive the broadcasted signal $\mathbf{y}^d$ though SD link as (\ref{SR-SD}).
    \STATE Predict the class $z^d$ through both $\mathbf{y}^f$ and $\mathbf{y}^d$ by the classifier at the destination node.
    \STATE Reconstruct $\mathbf{\hat{s}}^d$ by the class-aided JSCC decoder through both $\mathbf{y}^f$ and $\mathbf{y}^d$ with the knowledge of the true label $z$.
    \STATE Compute $L_3$ and update the parameters of the relay node's encoder, all modules in the destination node.
    \ENDWHILE
	\end{algorithmic}
\end{algorithm}
\subsection{Training algorithm}
\addtolength{\topmargin}{0.015in}
In the new architecture, the entire training algorithm is divided into three stages. In the first stage, the JSCC encoder at the source node and the JSCC decoder at the relay node are trained. In the second stage, the classifier at the relay node is trained. In the third stage, the class-aided JSCC encoder at the relay node, the class-aided JSCC decoder at the destination node, the classifier, and the mutual attention module are jointly trained.


During the first stage, the mean squared error (MSE) is used as the loss function, based on the input souce $\mathbf{s}$ and the reconstruction $\mathbf{\hat{s}^r}$. the loss function can be expressed as 
\begin{align}
\mathcal{L}_{1}(\mathbf{s},\mathbf{\hat{s}}^r) = \mathbb{E}_{\mathbf{s}\sim p(\mathbf{s})} ||\mathbf{s}-\mathbf{\hat{s}}^r||_2^2,
\end{align}
where $p(\mathbf{s})$ represents the distribution of $\mathbf{s}$. 
The objective is to enable the relay node to recover the original image through semantic communications.

In the second stage, the classifier at the relay node is trained independently with the reconstructed image as input to avoid influence the performance of reconstruction arising from the balance of the two tasks. 
When training the classifier, cross-entropy is used as the performance metric, and the loss function can be formulated as
\begin{align}
\mathcal{L}_{2}(p_{\mathbf{\hat{s}}^r}(z^r), q_{\mathbf{\hat{s}}^r}(z^r)) = -\mathbb{E}_{\mathbf{\hat{s}}^r\sim p(\mathbf{\hat{s}}^r)}[p_{\mathbf{\hat{s}}^r}(z^r) \log  q_{\mathbf{\hat{s}}^r}(z^r)],
\end{align}
where ${p_{\mathbf{\hat{s}}^r}(z^r)}$ denotes the true probability distribution of the class label of the images in the distribution $p(\mathbf{\hat{s}}^r)$, and ${q_{\mathbf{\hat{s}}^r}(z^r)}$ represents the predicted probability distribution by the classifier in the relay node.

The third stage involves multi-task joint training, which includes both the recovery of the original image and the target classification. Therefore, our performance metric is the weighted sum of MSE and cross-entropy, and the loss function can be formulated as
\begin{align}
&\mathcal{L}_{3}(\mathbf{s}, \mathbf{\hat{s}}^d, p_{\mathbf{y}^d,\mathbf{y}^f}(z^d), q_{\mathbf{y}^d,\mathbf{y}^f}(z^d))=\nonumber\\
&\mathcal{L}_{1}(\mathbf{s}, \mathbf{\hat{s}}^d) + \lambda\mathcal{L}_{2}(p_{\mathbf{y}^d,\mathbf{y}^f}(z^d), q_{\mathbf{y}^d,\mathbf{y}^f}(z^d)),
\end{align}
where $\lambda$ represents the weight assigned to the classification task.
It should be noted that during the training process, the class label input to the class-aided JSCC is the true label of the image, and after training, the predicted labels output by the classifiers are used for transmission. This is because the predictions may be incorrect, and during the training process, it is necessary to ensure that the class-aided JSCC has the correct information about the true label to learn the correct semantics for better performance. In practice, the true label is difficult to obtain, so only the predicted labels can be input into the class-aided JSCC. The training algorithm is summarized in Algorithm \ref{trainCDESC}.

\section{EXPERIMENTS RESULTS}
\begin{figure}[t]
  \begin{center}
    \includegraphics[width=0.41\textwidth]{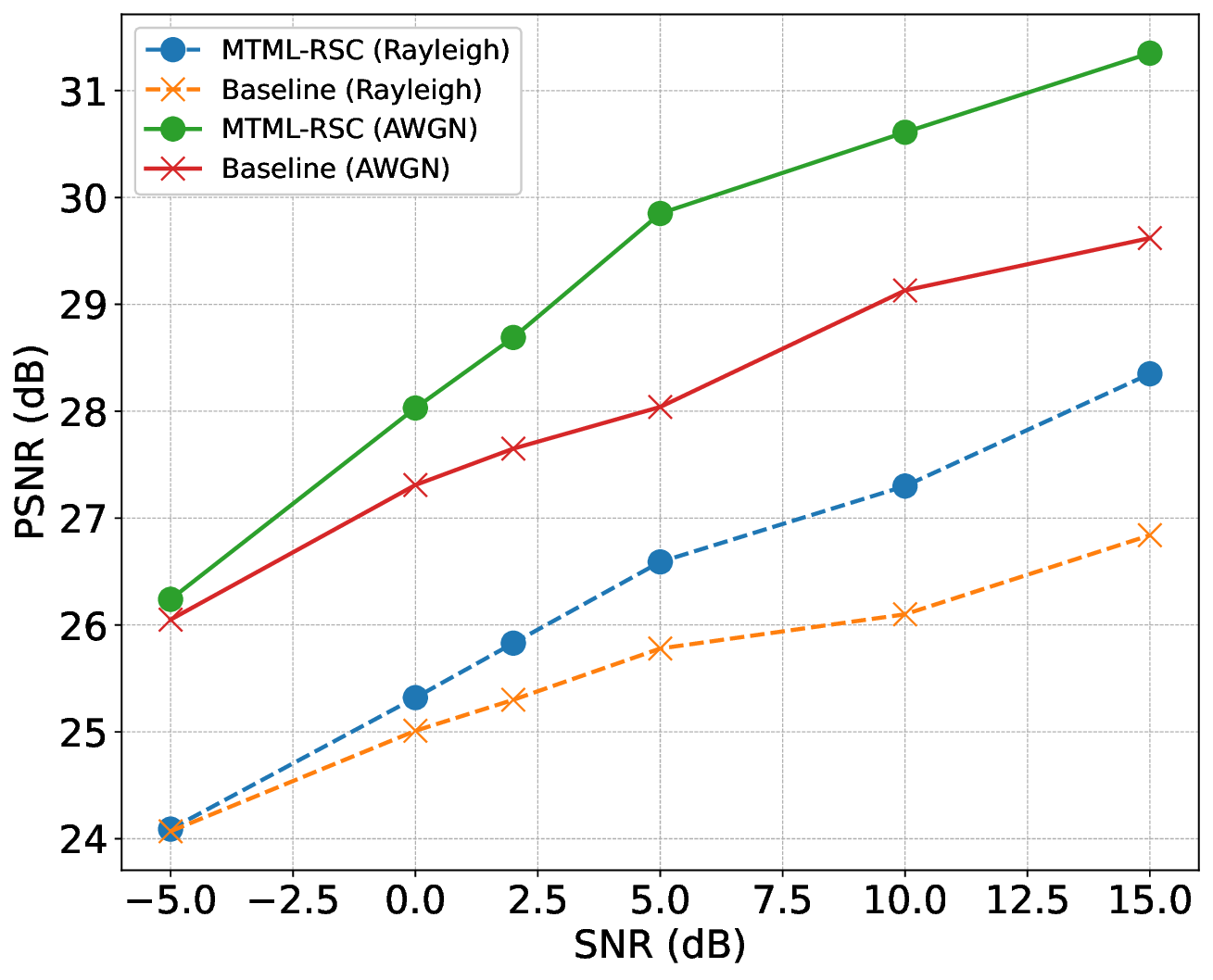}
  \end{center}
    \caption{The PSNR performance versus SNR with $d_{SR}=d_{RD}=0.5$. Each link are both the AWGN and Rayleigh fading channels.}
    \label{psnr} 
    \vspace{-0.5 cm}
\end{figure}

In this section, we present the detailed experimental setup and results to validate the effectiveness of our scheme. 

\subsection{Experiment Setup}
\begin{figure}[t]
  \begin{center}
    \includegraphics[width=0.42\textwidth]{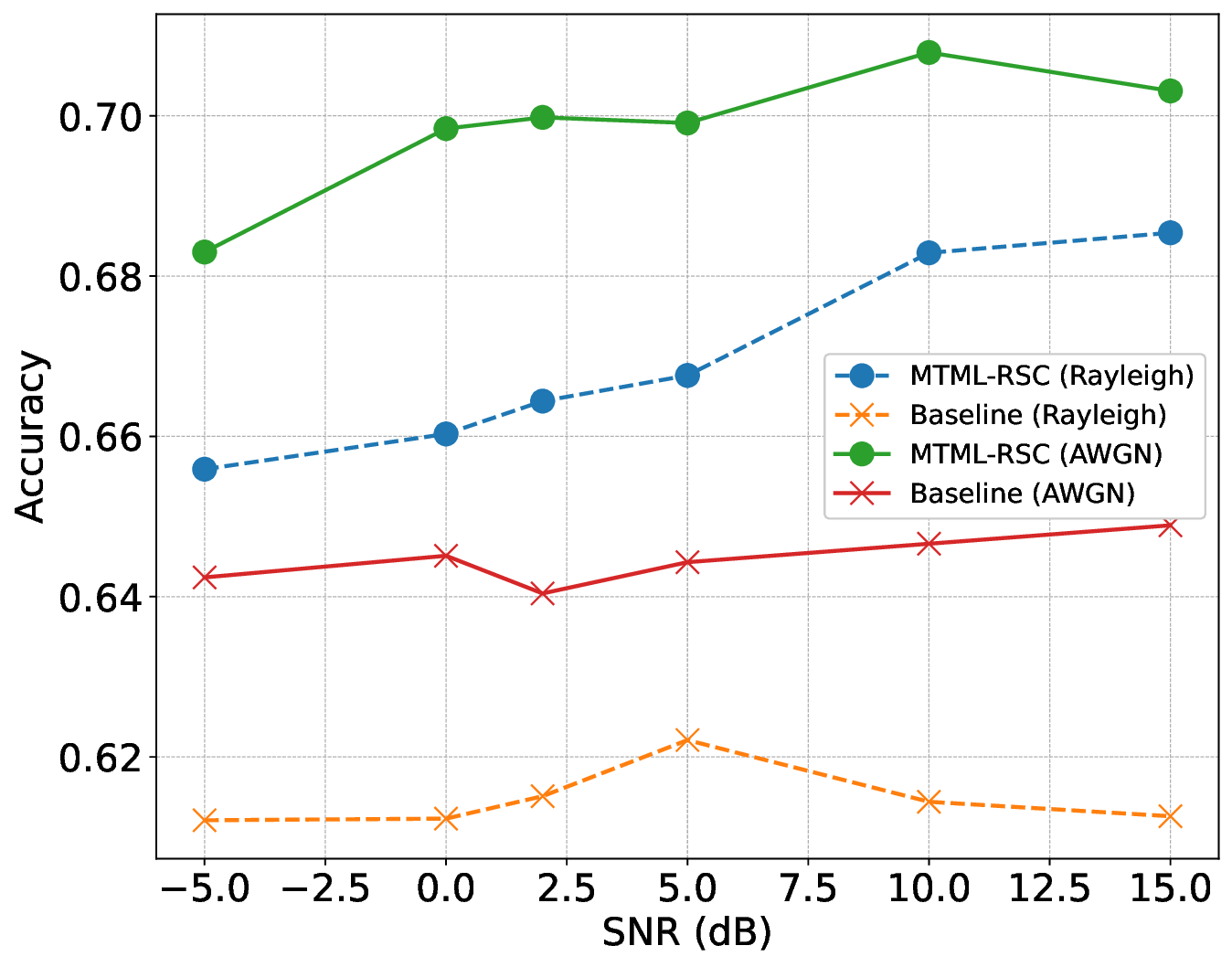}
  \end{center}
    \caption{The accuracy versus SNR with $d_{SR}=d_{RD}=0.5$. Each link are both the AWGN and Rayleigh fading channels}
    \label{acc} 
    \vspace{-0.5 cm}
\end{figure}
In our experiments, we adopt the STL10 dataset with 10 classes, where each class contains 500 images for training and 800 images for evaluation. The images come from the famous and widely adopted ImageNet dataset with size $96\times 96$.
For the position of the three nodes, we adopt a normalized distance with $d_{SD}=1$ and $d_{SR}+d_{RD}=1$ with their values in the range from $0$ to $1$. Also, we control $P_s=P_r$ and set $\sigma_{SR}=\sigma_{RD}=\sigma_{SD}$ for clarity. In this way, the signal-to-noise ratio (SNR) is defined as 
$SNR=10\log_{10}\left(\frac{P_s}{\sigma^2_{SD}}\right)$.

 Additionally, we employ a minimum mean square error (MMSE) equalizer to process the signals transmitted over a Rayleigh fading channel.
We take the scheme with only the relay link as the baseline, which has the same architecture as our MTML-RSC except the mutual attention module at the destination node and the classifier at the relay node. The class-aided JSCC is also replace with the normal JSCC.
Both schemes are evaluated with peak-signal-to-noise ratio (PSNR) for reconstruction and accuracy for classification. 
The parameters of the models in both schemes are trained and evaluated individually at each different SNR and distance with Adam optimizer at the learning rate of $10^{-4}$ for over $400$ epoch for better performance.
For the basic configurations of MTML-RSC and baseline, the Swin Transformer block number is $[n_1^e,n_2^e]=[n_1^d,n_2^d]=[2, 4]$ and patch number is $[c_1^e,c_2^e]=[c_1^d,c_2^d]=[128, 256]$. The channel bandwidth ratio in all experiments is $\frac{1}{12}$.


\subsection{Result Analysis}
Fig. \ref{psnr} and \ref{acc} illustrate the performance of PSNR and accuracy versus SNR with all links to be AWGN and Rayleigh fading with $d_{SR}=d_{RD}=0.5$, respectively. It can be observed that the proposed MTMD-RSC achieves performance gain in both PSNR and accuracy simultaneously compared to the baseline. For example, at the SNR of $15$ dB, we achieve $1.73$ dB gain in PSNR with accuracy increasing from $64.89\%$ to $70.31\%$ under the AWGN channel while under the Rayleigh fading channel, $1.51$ dB gain in PSNR with accuracy increasing from $61.26\%$ to $68.54\%$ is achieved. 
Meanwhile, the performance gain is gradually reduced as the decrease of SNR, e.g. only $0.19$ dB gain in PSNR with $4.06$ percentage points in accuracy under the AWGN channel and $0.02$ dB gain in PSNR with accuracy from $61.23\%$ to $66.03\%$ under Rayleigh fading channel at the SNR of $-5$ dB. 
The phenomenon arises from the fact that as the SNR decreases, the information provided by the SD link is reduced. Therefore, the performance gain coming from combining the link is reduced.
These results demonstrate that the proposed MTMD-RSC can effectively combining the signals from the direct link to enhance the performance of image reconstruction and target classification simultaneously.

\begin{figure}[t]
  \begin{center}
    \includegraphics[width=0.435\textwidth]{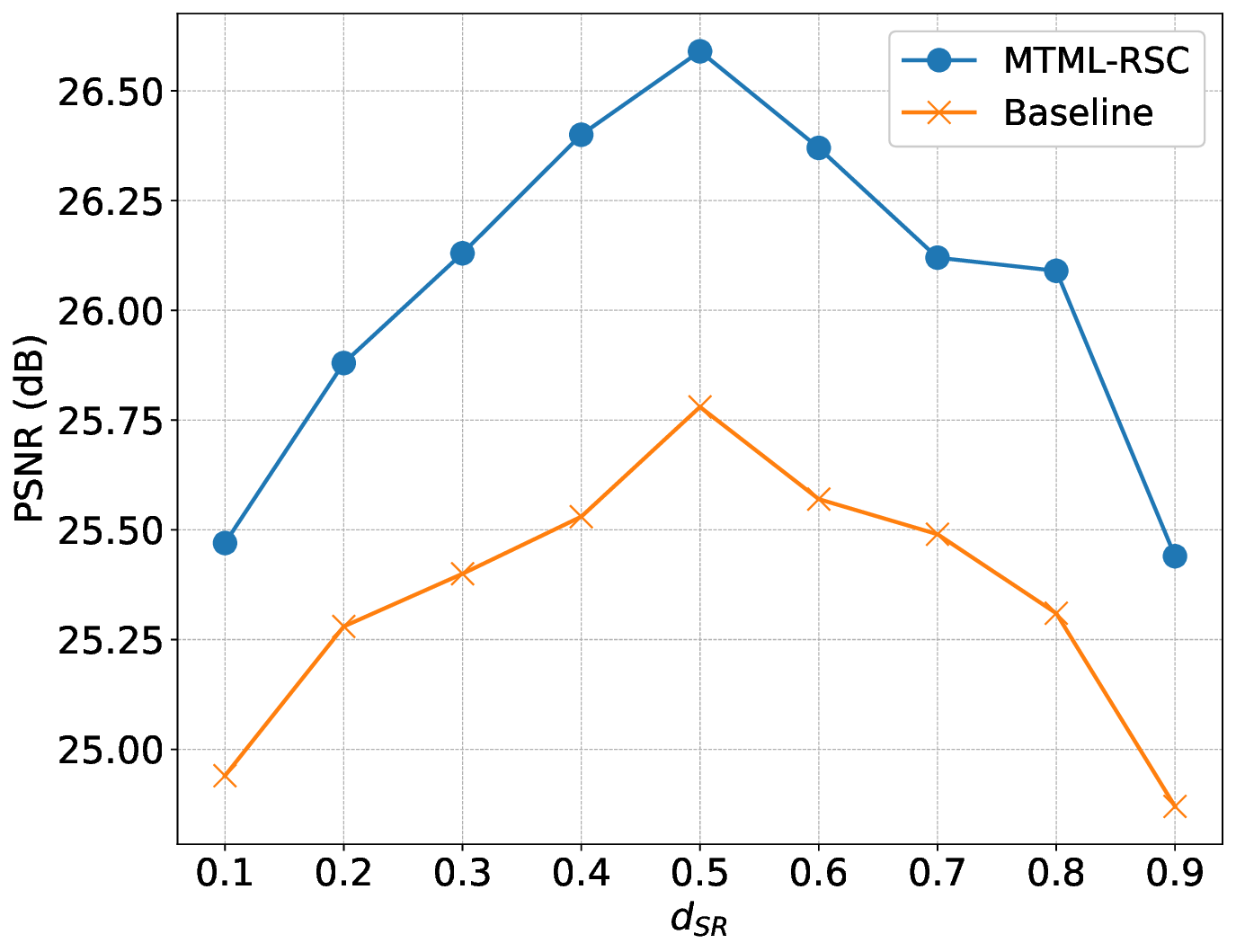}
  \end{center}
    \caption{The PSNR versus $d_{SR}$ with $SNR=5$ dB. Each link is the Rayleigh fading channel.}
    \label{dis_psnr} 
    \vspace{-0.5 cm}
\end{figure}

\begin{figure}[t]
  \begin{center}
    \includegraphics[width=0.43\textwidth]{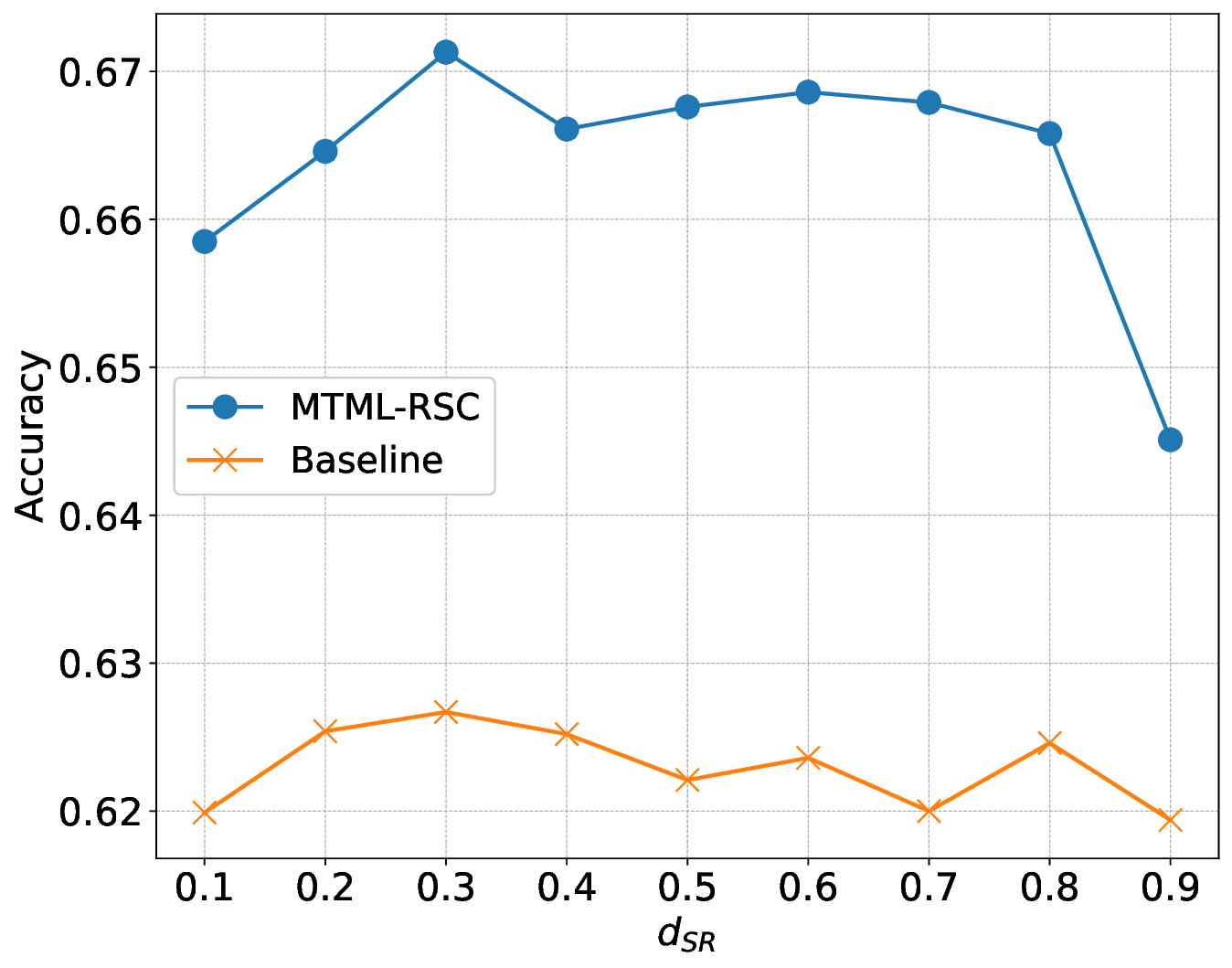}
  \end{center}
    \caption{The accuracy versus $d_{SR}$ with $SNR=5$ dB. Each link is the Rayleigh fading channel.}
    \label{dis_acc} 
    \vspace{-0.5 cm}
\end{figure}

Fig .\ref{dis_psnr} and \ref{dis_acc} show the variation in PSNR and accuracy as the relay node moves from near the source with $d_{SR}=0.1$ and $d_{RD}=0.9$, to near the destination with $d_{SR}=0.9$ and $d_{RD}=0.1$.
It is clear in PSNR that the schemes achieve optimal performance when the relay node is positioned at the midpoint, while gradually degraded as the relay moves to the two points. For example, as the $d_{SR}$ increase from $0.5$ to $0.9$, the PSNR decrease from $26.59$ dB to $25.44$ dB and the gain decrease from $0.81$ dB to $0.53$ dB.
The accuracy exhibits a general same trend of being lower at both ends and higher in the middle, with some fluctuations. which is arising from the balance of multi-tasks. 
The results illustrate that the relay node should be positioned at the midpoint to achieve optimal performance.

\section{CONCLUSION}
In this paper, we propose the MTML-RSC scheme for relay semantic communications to combine the direct link and the relay link for enhancing image reconstruction quality and classification accuracy simultaneously. We reveal that the two tasks can benefit each other and also classification requires significantly less information than reconstruction. Thereby, we proposed a semantic-focused forward, which utilizes the designed classifier-aided JSCC encoder to extract and transmit the semantics of the focused class. Furthermore, the destination node applies the proposed multi-link combining receiving, where the signals from both links are combined through the mutual attention module for classification, which then guides the focus of the class-aided decoder for reconstruction.
Experimental results illustrate that the proposed MTML-RSC achieves significant gains in PSNR and accuracy simultaneously, proving the effective utilization of the direct link.

\footnotesize
\bibliographystyle{IEEEtran}
\bibliography{reference}{}
\end{document}